\documentstyle[twoside,psfig]{article}
\input{ijmpb-macros}

\begin{document}

\runninghead{$s$- and $d$-wave Symmetries in Nonadiabatic Theory
of Superconductivity}{$s$- and $d$-wave Symmetries in Nonadiabatic Theory
of Superconductivity}

\normalsize\textlineskip
\thispagestyle{empty}
\setcounter{page}{1}

\copyrightheading{}			

\vspace*{0.88truein}

\fpage{1}
\centerline{\bf $s$- AND $d$-WAVE SYMMETRIES IN NONADIABATIC}
\vspace*{0.035truein}
\centerline{\bf THEORY OF SUPERCONDUCTIVITY}
\vspace*{0.37truein}

\centerline{\footnotesize PAOLA PACI}
\vspace*{0.015truein}
\centerline{\footnotesize\it Dipartimento di Fisica, 
Universit\`a ``La Sapienza'', P.le Aldo Moro 2}
\baselineskip=10pt
\centerline{\footnotesize\it Roma, 00185, Italy} 
\baselineskip=10pt
\centerline{\footnotesize\it and INFM, Unit\`a Roma1}

\vspace*{10pt}
\centerline{\footnotesize CLAUDIO GRIMALDI}
\vspace*{0.015truein}
\centerline{\footnotesize\it \'Ecole Polytechnique F\'ed\'erale, 
D\'epartment de microtechnique IPM}
\baselineskip=10pt
\centerline{\footnotesize\it Lausanne, CH-1015, Switzerland}

\vspace*{10pt}
\centerline{\footnotesize LUCIANO PIETRONERO}
\vspace*{0.015truein}
\centerline{\footnotesize\it Dipartimento di Fisica, 
Universit\`a ``La Sapienza'', P.le Aldo Moro 2}
\baselineskip=10pt
\centerline{\footnotesize\it Roma, 00185, Italy} 
\baselineskip=10pt
\centerline{\footnotesize\it and INFM, Unit\`a Roma1}

\vspace*{10pt}
\centerline{\normalsize and}

\vspace*{10pt}
\centerline{\footnotesize EMMANUELE CAPPELLUTI}
\vspace*{0.015truein}
\centerline{\footnotesize\it Dipartimento di Fisica, 
Universit\`a ``La Sapienza'', P.le Aldo Moro 2}
\baselineskip=10pt
\centerline{\footnotesize\it Roma, 00185, Italy} 
\baselineskip=10pt
\centerline{\footnotesize\it and INFM, Unit\`a Roma1}
\vspace*{10pt}

\abstracts{High-$T_c$ superconductors 
have Fermi energies $E_F$ much smaller than conventional
metals comparable to phonon frequencies.
In such a situation nonadiabatic effects are important.
A generalization of Eliashberg theory in the nonadiabatic
regime has previously been shown to reproduce some anomalous features of the
high-$T_c$ superconductors as for istance the enhancement  
of $T_c$ or the isotopic effects on $T_c$ and $m^*$.
In this contribution we address 
the issue of the symmetry of the gap in the context of nonadiabatic
superconductivity.
We show that vertex corrections have a momentum structure  
which favours $d$-wave superconductivity when forward scattering is
predominant. An additional increase of $T_c$ is also found.}{}{}

\textlineskip			
\vspace*{12pt}			

\textheight=7.8truein
\setcounter{footnote}{0}
\renewcommand{\thefootnote}{\alph{footnote}}

\section{Introduction}

High temperature superconductors (HTSC) show many ``anomalous'' features
which are not reported in conventional materials.
One of these peculiarities is the strong dependence of the order parameter
on the momentum ${\bf k}$ : $\Delta=\Delta({\bf k})$.
Josephson tunneling experiments and photoemission spectroscopy
show that the superconducting gap of many cuprates has a 
dominant $d$-wave symmetry: $\Delta({\bf 
k})\simeq\Delta[\cos(k_x)-\cos(k_y)]$\cite{shen}.

The origin of the $d$-wave symmetry in HTSC is still controversial.
On one hand, the observed $d$-wave symmetry is regarded 
as an evidence against any purely electron-phonon pairing interaction
so that the mechanism 
responsible for superconductivity should be sought among pairing mediators
of electronic origin (like antiferromagnetic
fluctuations) with eventually a minor electron-phonon component. 
On the other hand, several theoretical studies have shown
that the {\em el-ph} interaction could give arise, under some
quite general circumstances, a $d$-wave symmetry of the condensate.
This could happen when for example charge carriers experience an
on-site repulsive interaction together with a phonon induced attraction
for large inter-electrons distances. The on-site repulsion
inhibits the isotropic $s$-wave superconducting response leading
the system to prefer order parameters of higher angular momenta.

Another important feature of the HTSC compunds is the low density
of the charge carriers\cite{uemura}. As a consequence, the Fermi energy $E_F$
in these materials is extremely small and can be comparable with
the phononic frequency scale $\omega_{ph}$. In such a situation one of the
assumptions
of the conventional theory of superconductivity, Migdal's theorem,
does not hold and a generalization of the standard equation in nonadiabatic
regime ($\omega_{ph}\simeq E_F$) is necessary.

The small density of charge carriers moreover implies
a reduced electronic screening of eventual charge fluctuations
like as the ones induced by {\em el-ph} interaction.
This leads to a momentum modulation of the {\em el-ph} interaction:
the long range contribution (small momenta) is characterized  by the 
unscreened {\em el-ph} interaction and is therefore attractive while
the local scattering at short range (large momenta)
is mainly dominated by the stong electronic correlation
and is repulsive. This structure of the interaction
favours a $d$-wave symmetry of the order parameter.

The experimental evidence of a $d$-wave component of the superconducting
order parameter and the breakdown of Migdal's theorem
are thus two unavoidable  aspects to be taken into account.

\section{The Model}

\begin{figure}[t]
\centerline{\psfig{figure=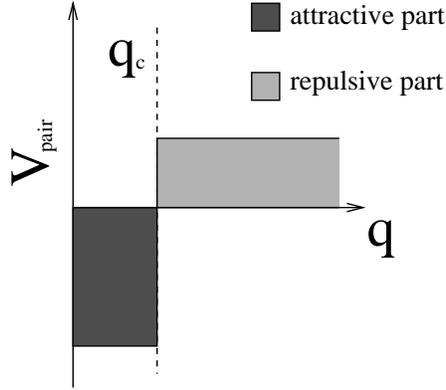,width=6cm,clip=!}}
\caption{Sketch of the total ({\em el-ph} + Coulomb) interaction
in momentum space.}
\label{f1}
\end{figure}

In this section we introduce a simple model interaction suitable
for our investigation beyond Migdal's limit and capable of
providing for $s$- or $d$-wave symmetries of the order parameter.
To define the model interaction $V_{\rm pair}(k-k')$ we have made use 
of a number of informations gathered from previous studies. 
First, in order to obtain order
parameters with higher angular momenta than $s$-wave, it is
sufficient to consider a pair interaction made of a repulsive
part at short distances and an attractive one at higher distances.
In momentum space, this interaction corresponds to an attractive
coupling for small ${\bf q}$ and a repulsive one for large ${\bf q}$,
where ${\bf q}={\bf k}-{\bf k}'$ is the momentum transfer
(Fig. \ref{f1}):
\begin{equation}
V_{\rm pair}({\bf q},\omega_m)=
\left[-|g({\bf q})|^2\theta(q_{\rm c}-|{\bf q}|)
+U\theta(|{\bf q}|-q_{\rm c})\right]
D(\omega_m)\,,
\label{pair} 
\end{equation}
where $D$ represents the retarded interaction.

Let us now try to interpret this strong momentum modulation
in terms of $e$-$ph$ and electron-electron interactions.
In strongly correlated systems, as HTSC materials, 
the {\em el-ph} interaction
acquires an important momentum dependence in such a way that
for large values of the momentum transfer
${\bf q}$ the {\em el-ph} interaction is suppressed, whereas
for small values of ${\bf q}$ it is enhanced \cite{zey}.
A physical picture to justify this momentum modulation is 
the following \cite{dany}.
In many-electron systems a single charge carrier
is surrounded by its own correlation hole of size $\xi$
which can be much larger than the 
lattice parameter $a$ in the strongly correlated regime. 
This means that any electron interacts only with lattice distorsions
with wavelength larger than $\xi$, leading to
an effective upper cut-off $q_c\simeq\xi^{-1}$ in the momenta space.
Thus we have a non zero  electron-phonon interaction when 
$|{\bf q}|<q_c$.  The cut-off momentum
$q_c$ can also be regarded as a measure of the correlation in the system: 
$aq_c\ll 1$ in strongly correlated 
systems while $aq_c\simeq 1$ in the case of free electrons.
From the above consideration, the attractive
part at small ${\bf q}$ of our model pairing interaction
(the first term of Eq. (\ref{pair})) finds a natural interpretation in 
{\em el-ph} coupling modified by the strong electron correlations.

Having defined the nature of the attractive part of the total
pairing interaction,
we offer now a possible interpretation
for the remaining repulsive part acting at large ${\bf q}$
(second term in Eq. (\ref{pair})).
This repulsion is given by the residual {\it e}-{\it e} 
interaction and its momentum dependence can be obtained, in
analogy with the renormalization of the {\it e-ph} interaction,
by using the above picture of correlation holes.
In this picture, the residual {\it e}-{\it e}  interaction
represents the impossibility to overlap two correlation holes. 

\section{d-Wave Nonadiabatic Superconductivity}

The breakdown of Migdal's theorem in high-$T_{\rm c}$ 
superconductors inevitably calls for a generalization beyond the 
Migdal-Eliashberg (ME) scheme 
to include the no longer negligible vertex corrections.
A possible way to accomplish this goal is to rely on a perturbative
scheme by truncating the infinite set of vertex corrections
at a given order. In previous works, we have proposed 
a perturbative scheme in which the role of small parameter
is played roughly by $\lambda\omega_{ph}/E_{\rm F}$ 
($\lambda$ is the {\em el-ph} coupling) leading to
a generalized ME theory which includes the first
nonadiabatic vertex corrections \cite{GPSprl,PSG}. 
A diagramatic representation of the effective interaction (i.e. with the 
inclusion of vertex corrections) 
in the Cooper's channel
is shown in Fig. \ref{f2}, where we only include the 
terms that give a finite contribution for $T=T_c$.

A sistematic study of the several gap symmetries involves the projection 
of the order parameter on the different harmonic functions:  
\begin{figure}[t!]
\centerline{\psfig{figure=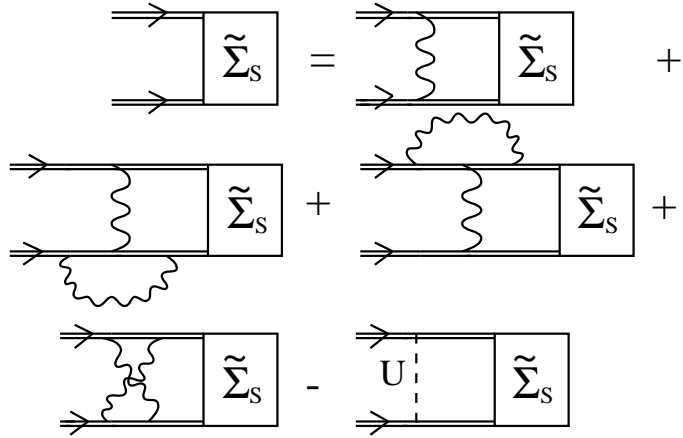,width=9cm}}
\caption{Self-consistent gap equation including the first corrections 
beyond Migdal's theorem}
\label{f2}
\end{figure}
\begin{equation}
\Delta(\phi)=
\sum_{l=-\infty}^{+\infty}\Delta^{(l)}Y_l(\phi)\ ,
\end{equation}
where $\phi$ is the polar angle on the Fermi's surface 
and $Y_l(\phi)=e^{il\phi}/\sqrt{2\pi}$ are eigenfunctions
of the operator $L=-id/d\phi$. In this paper we focus only on the interplay
between $s$- and $d$-wave symmetries.

A purely $s$-wave contribution is recovered when interaction is totaly 
attractive, namely for $q_c\leq 2k_F$. In the limit $q_c\rightarrow 0$
the $s$-wave component is suppressed and only the $d$ symmetry survives.
From qualitative point of view we expect a crossover from $s$-wave
superconductivity to  $d$-wave superconductivity as function of the parameter
$q_c$ and therefore as function of the degree of electronic correlation
of the system.

\begin{figure}
\centerline{\psfig{figure=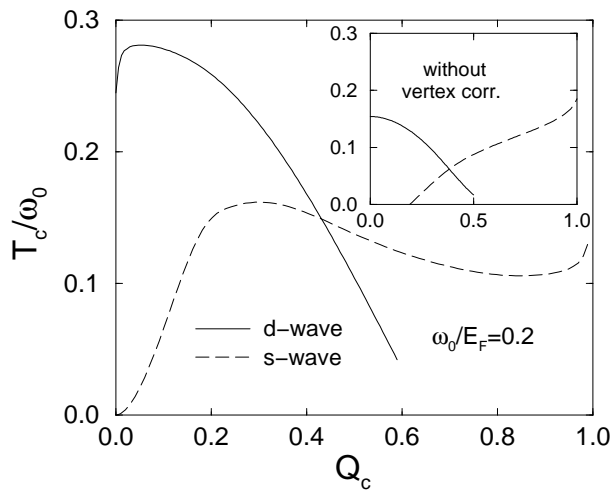,width=8cm}}
\caption{Behaviour of the critical temperature $T_c$
as function of $Q_c$ in the $s$- and $d$-wave 
symmetry channels. 
The case shown refers to the parameters $\lambda=1$ and 
$\mu^*=0.1$. In the inset
it is shown the case without nonadiabatic corrections.} 
\label{tc}
\end{figure}

The generalized ME equations including the nonadiabatic corrections
are numerically solved to calculate the critical temperature $T_c$
as function of the dimensionless parameter $Q_c=q_c/2k_F$\cite{mio}.
In Fig. \ref{tc} we show the dependence of $T_c$ on $Q_c$ in 
the weak repulsion case
$U/g^2=0.1$ for an Einstein phonon spectrum with frequency $\omega_0=0.2E_F$.
The inclusion of the nonadiabatic contributions produces a 
marked enhancement
of the critical temperature for small values of $Q_c$, where vertex corrections
yield an increase of the effective electron-phonon interaction\cite{GPSprl,PSG}.

\newpage
\section{Conclusions}

We have studied a model
interaction in which the {\em el-ph} interaction is dominant at small values
of ${\bf q}$ and the residual repulsion of electronic origin is instead
important at larger momentum transfers, we have shown that also when
the solution has $d$-wave symmetry, the inclusion of nonadiabatic corrections
enhances $T_c$ compared to the case without corrections.
This result is a consequence of the weak dependence at small ${\bf q}$
on the particular symmetry averages, $s$- or $d$-wave, of the vertex
and cross corrections.

It is finally interesting to compare the present results
with the phenomenology of the oxide SC, which show $d$-wave,
and the fullerides, which instead show $s$-wave.
In our perspective there are important differencies
between the two materials. The main one is that the oxides
have their largest values of $T_c$ when the Fermi surface
is strongly influenced by Van Hove singularities. Then
correlation effects can be estimated to be larger in the oxides
and, finally, fullerides seem to have rotational disorder
which wuold favour $s$-wave.
Therefore, in principle, it could happen that in the oxides,
going into the overdoped phase might lead to a crossover
from $d$-wave to $s$-wave depending on the parameters.
Also in this perspective one can understand the different behavior
of the Nd electron doped compounds which seem to show
$s$-wave. We conjecture that this is due to the possibly
large distance of the Fermi surface with the respect to the Van Hove
singularity.

\end{document}